\documentstyle[PASJadd]{PASJ95}
\draft
\begin{document}

\title{Evolution of Hydromagnetic Disturbances
in Low Ionized Cosmic Plasmas}
\author{Hideko {\sc Nomura},$^1$ Hideyuki {\sc Kamaya},$^2$
and Shin {\sc Mineshige}$^1$
\\[12pt]
$^1${\it Department of Astronomy, Faculty of Science, Kyoto University,
Sakyo-ku, Kyoto 606-8502}\\
{\it E-mail(HN): nomura@kusastro.kyoto-u.ac.jp}\\
$^2${\it Department of Physics, Faculty of Science, Kyoto University,
Sakyo-ku, Kyoto 606-8502}}

\abst{We consider the propagation of hydromagnetic waves
generated by a compact turbulent source in low ionized plasmas,
applying the Lighthill theory.
We assume the plasma to be isothermal, and adopt
a uniform, stationary medium
thread by ordered magnetic fields as an initial condition.
Then, the distinct properties of the hydromagnetic waves
originating from a source oscillating with a fixed frequency 
are studied in the linear regime. As is well known, in low ionized plasmas, 
the generated waves dissipate due to ion--neutral damping.
In this paper, the dependence of the dissipation rate
on the frequency of the oscillating source is investigated.
The larger the frequency becomes, the more substantial
is the wave dissipation. 
Implications of our results on the energy source in molecular clouds 
are also discussed. Interestingly, since 
the outflow lobes associated with young stellar objects act
as compact turbulent sources, 
hydromagnetic waves are generated by them.
{}From our order-estimations, 
about 70\% of the energy of the outflow itself propagates
as waves or turbulences, while the remaining 30\% dissipates 
and heats the neutrals via ion--neutral damping.
Then, we confirm that the outflows are significant energy sources
in molecular clouds in the context of the Lighthill theory.}
\kword{Bipolar flows --- Interstellar: clouds ---
Interstellar: magnetic fields --- Magnetohydrodynamics --- Wave}

\maketitle
\thispagestyle{headings}

\section{Introduction}

Young stellar objects (YSOs) are believed to be formed within 
dense clumps in molecular clouds (Myers, Benson 1983).
Going with the gas condensation associated with star formation, 
a large amount of gravitational energy must be released,
and a part of the released energy drives the jets and/or outflows
(e.g., Shu et al.\ 1993; Kudoh, Shibata 1997; Tomisaka 1998).
In this paper, we mainly discuss the role of the outflows.
The total mechanical luminosities of all the outflows 
in a molecular cloud are observed, for example,
to be $\sim$ 33 $\LO$ in Orion A (Fukui et al.\ 1993)
and $\sim$ 43.5 $\LO$ in Taurus (Bontemps et al.\ 1997).
We stress the importance of the activity of the outflows
on the ambient molecular gas.

Lighthill (1952) discussed the aerodynamical generation of waves
from a compact turbulent source; his theory was developed concerning
the generation of hydromagnetic waves in a magnetized medium 
(Kulsrud 1955; Kato 1967).
Waves originate from the pressure fluctuations 
caused by fluctuations in the momentum of velocity and
magnetic fields of unsteady turbulent flows.
Their theories have been applied to studies of
the heating mechanism of stellar chromosphere and coronae
(see Lee 1993 and references therein).
In this paper, we apply them to a study of molecular clouds,
where the ionization fraction is low.
In star--forming regions, the outflow lobe filled with turbulent molecular gas
(e.g., Stahler 1994) is one of the good candidates for a compact turbulent
source (e.g., Nomura et al.\ 1998). 
We then examine the potentiality of outflows as an energy source 
in molecular clouds in the context of the Lighthill theory.
Part of the energy of the outflows dissipates in the molecular gas
via the so-called ion--neutral damping
(Kulsrud, Pearce 1969; Zweibel, Josafatsson 1983;
Nakano 1998, see also Kamaya, Nishi 1998),
as far as the outflows can generate the hydromagnetic waves.

The problem concerning the heating source in molecular clouds 
is of interest. In molecular clouds (with a temperature of $\sim$ 10 K
and a number density of $\sim$ 10$^{3}$ cm$^{-3}$) the cooling processes
have been well studied (e.g., Goldsmith, Langer 1978; 
Neufeld et al.\ 1995). The heating processes, however, are
problematic because it is uncertain what is the dominant process.
Although it is often assumed that the main heating source is cosmic rays,
the heating rate due to H$_{2}$ formation on grains,
gas--grain collisions, compressions or collapses of dense cores, 
and turbulence may be comparable to that caused by cosmic rays
(Lizano, Shu 1987; Hollenbach 1988; Stone et al.\ 1998).
In this paper, we insist that the ion--neutral damping of waves
generated from outflows is also a dominant heating mechanism
(cf. Zweibel, McKee 1995).

We also discuss the role of outflows as an energy source to support
the turbulence in molecular clouds.
In many molecular clouds and cloud cores,
the scaling law of the line width--size relation, 
$\Delta v\propto L^{1/2}$, has been observed 
with various molecular lines
(Larson 1981; Falgarone et al. 1992; Caselli, Myers 1995).
The observed widths are wider than the thermal width and
the power-law index of the scaling law is similar to that of the Kolmogorov
turbulence. These facts suggest that the interstellar medium may be turbulent
(Larson 1981; Scalo 1987; Franco, Carrami\~{n}ana 1998).
If so, it should quickly dissipate
because of the shock, etc. (e.g., Stone et al.\ 1998);
thus, some kinds of energy sources are needed.
Many candidates have been suggested, e.g., galactic differential rotations,
isolated supernovae, stellar winds,
Alfv\'{e}n waves from collapsing cloud cores, expanding H\,{\footnotesize II}
regions, and bipolar outflows
(Norman, Silk 1980; Mouschovias, Paleologou 1980; Fleck 1981; 
Scalo 1987; Miesch, Bally 1994).
In recent numerical simulations 
it is suggested that the sources with stellar sizes may be 
more important than larger-scale sources 
(Gammie, Ostriker 1996; V\'{a}zquez-Semadeni et al.\ 1997).
In the following discussions, 
we estimate how much energy is given to the turbulence
through waves generated by outflows associated with YSOs.

In subsequent sections,
we investigate the evolution of hydromagnetic waves
generated from a compact turbulent source in a low--ionized, 
magnetized medium. We note that the difference between ours and
previous researches on waves in such medium
(e.g., Balsara 1996) is that we treat not only
the propagation of waves, but also their generation from a source.
Our formula could be extended to deal with problems concerning
the global evolution of molecular clouds.
For the first step, we present in this paper an exact solution
of our formula in the case that the source is modeled by
an oscillating source. We then apply it to suggest 
the role of outflows as a heating source and
an energy source of turbulence in molecular clouds.
In the next section 
we derive the basic equations and inhomogeneous wave equation.
In section 3 we estimate the energy of the waves from a magnetized turbulent
source.
Applications to molecular clouds are presented in section 4, and
finally we summarize the results in section 5.

\section{Basic Equations}

We consider the following two-fluid equations for a partially ionized,
magnetized, and compressible medium:
the continuity equation, the momentum equation, and the equation of state
for the neutrals,
\begin{equation}
\frac{\displaystyle{\partial\rho}}{\displaystyle{\partial t}}+\nabla\cdot(\rho\mbox{\boldmath $v$})=0,
\end{equation}
\begin{equation}
\rho\frac{\displaystyle{\partial\mbox{\boldmath $v$}}}{\displaystyle{\partial t}}+\rho(\mbox{\boldmath $v$}\cdot\nabla)\mbox{\boldmath $v$}=-\nabla p+\mbox{\boldmath $F$}_{\rm fric},
\end{equation}
and
\begin{equation}
p=c_{\rm s}^2\rho;
\end{equation}
the momentum equation for the ions,
\begin{equation}
\rho_{\rm i}\frac{\displaystyle{\partial\mbox{\boldmath $v$}_{\rm i}}}{\displaystyle{\partial t}}+\rho_{\rm i}(\mbox{\boldmath $v$}_{\rm i}\cdot\nabla)\mbox{\boldmath $v$}_{\rm i}=
-\nabla p_{\rm i}
+\frac{\displaystyle{1}}{\displaystyle{4\pi}}(\nabla\times\mbox{\boldmath $B$})\times\mbox{\boldmath $B$}-\mbox{\boldmath $F$}_{\rm fric},
\end{equation}
the induction equation,
\begin{equation}
\frac{\displaystyle{\partial\mbox{\boldmath $B$}}}{\displaystyle{\partial t}}=\nabla\times (\mbox{\boldmath $v$}_{\rm i}\times\mbox{\boldmath $B$}),
\end{equation}
and the divergence free condition of the magnetic field,
\begin{equation}
\nabla\cdot\mbox{\boldmath $B$}=0.
\end{equation}

In the above equations $\rho$, $\mbox{\boldmath $v$}$, $p$, $\mbox{\boldmath $B$}$, 
and $c_{\rm s}$ denote the density, velocity, pressure, 
magnetic field, and sound speed, respectively.
Subscript i is used to represent ions.
The force $\mbox{\boldmath $F$}_{\rm fric}$ is the frictional force given by
\begin{equation}
\mbox{\boldmath $F$}_{\rm fric}=\gamma\rho\rho_{\rm i}(\mbox{\boldmath $v$}_{\rm i}-\mbox{\boldmath $v$}),
\end{equation}
which arises from collisions between the ions and the neutrals.
Here, we approximate the drag coefficient by
$\gamma\approx 3.5 \times 10^{13}$ cm$^3$ g$^{-1}$ s$^{-1}$ 
(see e.g., Shu 1983).

We now consider cool molecular clouds where the ionization fraction 
is very small (the ratio of the number density of ions to that of neutrals,
$n_{\rm i}/n\approx 10^{-8}$ -- $10^{-6}$) and the collision time is short
compared with other concerned timescales.
In this condition,
we consider only the last two terms in equation (4),
which are the magnetic force and frictional force,
since the inertia of the ions to their motions is negligible.
We, therefore, use the following equation instead of the equation (4):
\begin{equation}
\frac{\displaystyle{1}}{\displaystyle{4\pi}}(\nabla\times\mbox{\boldmath $B$})\times\mbox{\boldmath $B$}\approx\gamma\rho\rho_{\rm i}(\mbox{\boldmath $v$}_{\rm i}-\mbox{\boldmath $v$}).
\end{equation}

We assume that the medium is initially uniform and stationary 
($\rho=\rho_0\equiv$ const., $\mbox{\boldmath $v$}=\mbox{\boldmath $v$}_0\equiv \mbox{\boldmath $0$}$, and
$\mbox{\boldmath $B$}=\mbox{\boldmath $B$}_0=(B_0, 0, 0)$, where $B_0\equiv$ const.), 
and then waves ($\rho=\rho_1$, $\mbox{\boldmath $v$}=\mbox{\boldmath $v$}_1$, and 
$\mbox{\boldmath $B$}=\mbox{\boldmath $B$}_1$) are generated from a compact turbulent source. 
Thus, we write the density, for example, as a superimposion of 
the zeroth-order undisturbed, uniform part, $\rho_0$, and 
the first-order perturbed part, $\rho_1$:
\begin{equation} 
\rho=\rho_0+\rho_1.
\end{equation}
In addition, we assume the isothermal condition.

Making use of the $x$-component of the perturbed vorticity $\xi$ and 
the divergence of the perturbed velocity $\eta$ given by
\begin{equation}
\xi=\frac{\displaystyle{\partial v_{1z}}}{\displaystyle{\partial y}}-\frac{\displaystyle{\partial v_{1y}}}{\displaystyle{\partial z}}
\end{equation}
and
\begin{equation}
\eta=\frac{\displaystyle{\partial v_{1x}}}{\displaystyle{\partial x}}+\frac{\displaystyle{\partial v_{1y}}}{\displaystyle{\partial y}}+\frac{\displaystyle{\partial v_{1z}}}{\displaystyle{\partial z}},
\end{equation}
we can eliminate $\mbox{\boldmath $v$}_{\rm i}$, $\rho_1$, and $\mbox{\boldmath $B$}_1$ 
from equations (1) to (9). We thus obtain the following linearized
inhomogeneous wave equations with a radiation source $Q_{\xi}(x,t)$ and 
$Q_{\eta}(\mbox{\boldmath $r$},t)$ (see the Appendix):
\begin{equation} 
\biggl[\frac{\displaystyle{\partial^2}}{\displaystyle{\partial t^2}}-\frac{\displaystyle{c_{\rm A}^2}}{\displaystyle{\gamma\rho_{\rm i}}}\frac{\displaystyle{\partial^3}}{\displaystyle{\partial t\partial x^2}}-c_{\rm A}^2\frac{\displaystyle{\partial^2}}{\displaystyle{\partial x^2}}\biggr]\xi(x,t) = Q_{\xi}(x,t)
\end{equation}
and
\begin{equation} 
\biggl[\frac{\displaystyle{\partial^4}}{\displaystyle{\partial t^4}}-\frac{\displaystyle{c_{\rm A}^2}}{\displaystyle{\gamma\rho_{\rm i}}}\nabla^2\frac{\displaystyle{\partial^3}}{\displaystyle{\partial t^3}}-(c_{\rm A}^2+c_{\rm s}^2)\nabla^2\frac{\displaystyle{{\partial^2}}}{\displaystyle{\partial t^2}}+\frac{\displaystyle{c_{\rm A}^2c_{\rm s}^2}}{\displaystyle{\gamma\rho_{\rm i}}}\nabla^4\frac{\displaystyle{\partial}}{\displaystyle{\partial t}}+c_{\rm A}^2c_{\rm s}^2\nabla^2\frac{\displaystyle{\partial^2}}{\displaystyle{\partial x^2}}\biggr]\eta(\mbox{\boldmath $r$},t) = Q_{\eta}(\mbox{\boldmath $r$},t),
\end{equation}
where $c_{\rm A}\equiv B_0/(4\pi\rho_0)^{1/2}$ is the Alfv\'{e}n speed.
We note that the wave equation for $\zeta\equiv\partial v_{1x}/\partial x$
is written in the same form as equation (13) (see equation (A18) 
in the Appendix).
Equation (12) represents the wave equation for the Alfv\'{e}n waves
and both equations (13) and (A18) are for fast and slow waves.
{}From these three equations, we can derive $v_{1x}, v_{1y},$ and $v_{1z}$
(Lighthill 1960). The source terms of equations (12) and (13) are derived as
\begin{equation}
Q_{\xi}(x,t)=[\nabla\times\mbox{\boldmath $q$}]_x
\end{equation}
and
\begin{equation}
Q_{\eta}(\mbox{\boldmath $r$},t)=\frac{\displaystyle{\partial^2}}{\displaystyle{\partial t^2}}(\nabla\cdot\mbox{\boldmath $q$})-c_{\rm A}^2\nabla^2\frac{\displaystyle{\partial q_x}}{\displaystyle{\partial x}},
\end{equation}
where
\begin{equation}
\mbox{\boldmath $q$}(\mbox{\boldmath $r$},t)=-\frac{\displaystyle{\partial}}{\displaystyle{\partial t}}[(\mbox{\boldmath $v$}_{\rm t}\cdot\nabla)\mbox{\boldmath $v$}_{\rm t}-(\nabla\times\mbox{\boldmath $h$}_{\rm t})\times\mbox{\boldmath $h$}_{\rm t}]-c_{\rm A}\{\nabla\times[\nabla\times(\mbox{\boldmath $v$}_{\rm t}\times\mbox{\boldmath $h$}_{\rm t})]\}\times\mbox{\boldmath $e$}_x.
\end{equation}
Here, $\mbox{\boldmath $v$}_{\rm t}$ and 
$\mbox{\boldmath $h$}_{\rm t}\equiv\mbox{\boldmath $B$}_{\rm t}/(4\pi\rho_0)^{1/2}$ denote 
the velocity and the Alfv\'{e}n velocity in a turbulent source, respectively,
and $\mbox{\boldmath $e$}_x=(1,0,0)$ represents the direction of the uniform 
magnetic fields. We assume that the gas is fully ionized 
in the turbulent region in deriving the above source terms
(e.g., in outflow lobes the temperature of the gas is high enough
to be fully ionized). We note that $q$, which comes from the non-linear 
terms in the basic equations, represents the fluctuation of the momentum 
flow in the turbulent source (see Kulsrud 1955; Kato 1968).
In the case of a fully ionized plasma 
(where $\mbox{\boldmath $F$}_{\rm fric}=\mbox{\boldmath $0$}$), the wave equations become
\begin{equation}
\biggl[\frac{\displaystyle{\partial^2}}{\displaystyle{\partial t^2}}-c_{\rm A}^2\frac{\displaystyle{\partial^2}}{\displaystyle{\partial x^2}}\biggr]\xi(x,t) = Q_{\xi}(x,t)
\end{equation}
and
\begin{equation} 
\biggl[\frac{\displaystyle{\partial^4}}{\displaystyle{\partial t}}-(c_{\rm A}^2+c_{\rm s}^2)\nabla^2\frac{\displaystyle{\partial^2}}{\displaystyle{\partial t^2}}+c_{\rm A}^2c_{\rm s}^2\nabla^2\frac{\displaystyle{\partial^2}}{\displaystyle{\partial x^2}}\biggr]\eta(\mbox{\boldmath $r$},t) = Q_{\eta}(\mbox{\boldmath $r$},t)
\end{equation}
(see Kulsrud 1955; Lighthill 1960; Kato 1968).

We now assume that the oscillating source has a fixed frequency $\omega_0$;
that is, the source terms are expressed as
$Q_{\xi}(x,t)=Q_{\xi}(x)\exp(-i\omega_0t)$ and
$Q_{\eta}(\mbox{\boldmath $r$},t)=Q_{\eta}(\mbox{\boldmath $r$})\exp(-i\omega_0t)$.
In this case, from equations (12) and (13), $\xi$ and $\eta$ are written as
follows by means of Fourier transformation:
\begin{equation}
\xi(x,t)=\xi_0e^{-i\omega_0t}\int_{\infty}^{\infty}\frac{\displaystyle{\tilde{Q}_{\xi}(k_x)\exp(ik_xx)}}{\displaystyle{D_{\xi}(k_x,\omega_0)}}dk_x
\end{equation}
and
\begin{equation}
\eta(\mbox{\boldmath $r$},t)=\eta_0e^{-i\omega_0t}\int_{\infty}^{\infty}\frac{\displaystyle{\tilde{Q}_{\eta}(\mbox{\boldmath $k$})\exp(i\mbox{\boldmath $k$}\cdot\mbox{\boldmath $r$})}}{\displaystyle{D_{\eta}(\mbox{\boldmath $k$},\omega_0)}}d^3\mbox{\boldmath $k$},
\end{equation}
where
\begin{equation}
D_{\xi}(k_x,\omega)=\omega^2+i\frac{\displaystyle{c_{\rm A}^2}}{\displaystyle{\gamma\rho_{\rm i}}}\omega k_x^2-k_x^2
\end{equation}
and
\begin{equation}
D_{\eta}(\mbox{\boldmath $k$},\omega)=\omega^4+i\frac{\displaystyle{c_{\rm A}^2}}{\displaystyle{\gamma\rho_{\rm i}}}\omega^3k^2-(c_{\rm A}^2+c_{\rm s}^2)\omega^2k^2-i\frac{\displaystyle{c_{\rm A}^2c_{\rm s}^2}}{\displaystyle{\gamma\rho_{\rm i}}}\omega k^2+c_{\rm A}^2c_{\rm s}^2k^2k_x^2,
\end{equation}
and $\tilde{Q}_{\xi}(k_x)$ and $\tilde{Q}_{\eta}(\mbox{\boldmath $k$})$
are the Fourier transformations of $Q_{\xi}(x)$ and $Q_{\eta}(\mbox{\boldmath $r$})$, 
respectively.
Here, $D_{\xi}(k_x,w)=0$ represents the dispersion relation for
the Alfv\'{e}n waves and $D_{\eta}(\mbox{\boldmath $k$},w)=0$ is for the fast and
slow waves (cf. Balsara 1996).
Integrating equation (19) by the residual method, we can obtain $\xi$ as
\begin{equation}
\xi(x,t)=\xi_0e^{-i\omega_0t}\cdot2\pi i\sum_p\displaystyle\mathop{\mbox{Res}}_{k_x=k_{xp}}\frac{\displaystyle{\tilde{Q}_{\xi}(k_x)\exp(ik_xx)}}{\displaystyle{D_{\xi}(k_{x},\omega_0)}},
\end{equation}
where $k_{xp}$ represent poles of the function in the integral (19),
which satisfy $D_{\xi}(k_{xp},w_0)=0$. Now, only the wavenumber which satisfies
$\Re(k_{xp})>0$ is used, since we consider that the waves are going forward. 
We can obtain $\eta$ in the same manner
(see Morse, Feshbach 1953; Lighthill 1978). 

In figure 1,
we show the real part of (a) $\xi(x)\equiv e^{i\omega_0t}\xi(x,t)$ 
and (b) $\eta(r)\equiv e^{i\omega_0t}\eta(r,t)$ 
in the case that $\tilde{Q_{\xi}}(k_x)$ and $\tilde{Q_{\eta}}(\mbox{\boldmath $k$})$ 
are constant [which means that $Q_{\xi}(x)$ and $Q_{\eta}(\mbox{\boldmath $x$})$ are 
Dirac delta functions,
representing the limiting case that the size of the source $l\rightarrow 0$].
We calculated $\eta(r)$ in the plane $x=0$.
In the calculations we put $c_{\rm s}=0.2$ km s$^{-1}$ and 
$c_{\rm A}=0.45$ km s$^{-1}$
(which corresponds to $B_0=10\ \mu$G and $n_0=10^3$ cm$^{-3}$).
$\xi(x)$ and $\eta(r)$ were normalized by $\xi(0)$ and $\eta(0)$, 
respectively, and $x$ and $r$ were in the units of pc.
The dashed lines, solid lines, and dotted lines represent the waves
from the source with frequencies of
$\omega_0/\gamma\rho_{\rm i}=$ 0.5, 1.0, and 5.0, respectively. 
In order to obtain $k_{p}$ satisfying $D_{\eta}(k_p,\omega_0)=0$,
we used Mathematica.
We note that the spatial distribution of the amplitudes of the waves 
do not vary with time, since we now consider that the source oscillates 
permanently with a certain frequency. Then,
figure 1 indicates that when the frequency of the source is low,
the waves dissipate less. In contrast, the waves dissipate soon
if the frequency is higher than the collision frequency, $\gamma\rho_{\rm i}$
(see also the discussions for figure 3 in section 3).

\section{Energy of Generated Hydromagnetic Waves}

In this section we estimate the energy of hydromagnetic waves
generated from a compact turbulent source.
The energy of the waves is estimated as
\begin{equation}
E_{\rm wave}=\frac{\displaystyle{\rho_0}}{\displaystyle{2}}\int_V\biggl[v_1^2+\frac{\displaystyle{B_1^2}}{\displaystyle{4\pi\rho_0}}+c_{\rm s}^2\biggl(\frac{\displaystyle{\rho_1}}{\displaystyle{\rho_0}}\biggr)^2\biggr]dV,
\end{equation}
where $v_1^2=v_{1x}^2+v_{1y}^2+v_{1z}^2$ and 
$B_1^2=B_{1x}^2+B_{1y}^2+B_{1z}^2$.
Because the dependence of the variables on the distance from the source is
the same as $\eta$, we approximately calculated the energy as
$E\propto\int\eta^2dV$.
We put $dV\approx 4\pi r^2dr\sim 4\pi c_{\rm A}^3t^2dt$,
since the head of the wave propagates with about Alfv\'{e}n speed
(which means $r\sim c_{\rm A}t$).
Figure 2 shows the time evolution of the energy of waves generated 
from an oscillating source with a frequency $\omega_0/\gamma\rho_{\rm i} = 10^{-8}$ 
(bold solid line), 0.5 (dashed line), 1.0 (solid line), 
2.0 (dot-dashed line),
and 5.0(dotted line). The normalization of the energy, $E_0$, 
is the energy provided to the waves by the source during $10^5$ years.
Figure 2 suggests that the energy of the waves is proportional 
to the time if the frequency of the source is very small.
On the other hand, when the frequency is higher than the collision 
frequency of an ion with neutrals, the time evolution of the energy
has an upper limit.

These results can be understood from the dispersion relations
of the hydromagnetic waves.
In figure 3, we show the imaginary part of the wavenumber, $k_{\rm I}$, 
as a function of the frequency of the source, $\omega_0$.
The solid line, dashed line and dotted line denote the $k_{\rm I}$s of 
the Alfv\'{e}n waves, the fast waves, and the slow waves, respectively.
When the frequency approaches zero, $k_{\rm I}$ also becomes zero 
for all three waves. Since the waves are proportional to $\exp(ik_xx)$ or
$\exp(ikr)$, as can be seen in equation (23), $k_{\rm I}$ can represent 
the dissipation rate of the waves.
Thus, in the limit of vanishing frequency, 
there is no dissipation, as shown in figure 3. 
This leads to the strong conclusion that if the mechanical luminosity 
of a source has a constant value during its duration, 
the generated energy of the waves is nearly conserved; 
then, the generated total energy is proportional to that duration.
On the other hand, $k_{\rm I}$ becomes larger as the frequency increases,
which means that the waves soon dissipate.
Then, the energy scarcely increases over the time
when the head of the waves reach a certain radius from the source.
Thus, the upper limit appears as seen in figure 2.

In summary, the ratio of the dissipated energy to the energy of 
the propagating waves varys depending on the frequency of the source,
and evolve with the time. 
The ratio (dissipated/propagating) becomes larger as the frequency becomes
smaller or as time passes.

\section{Discussions}

In the previous section, we estimate the energy of the waves generated
from a compact turbulent source. Here, we apply the results
to the case in molecular clouds, 
where the outflow lobe associated with
a young stellar object is a good candidate for a compact turbulent
source (e.g., Nomura et al.\ 1998).

\subsection{Energy of Waves Generated from Outflows}

We first calculate the characteristic frequency and the energy of 
the outflow lobes 
in a molecular cloud, $\omega_{\rm flow}$ and $E_{\rm flow}$ as
\begin{equation}
\omega_{\rm flow}\sim\frac{\displaystyle{u_{\rm flow}}}{\displaystyle{l_{\rm flow}}}\sim 3\times 10^{-13}{\rm s}^{-1}\biggl(\frac{\displaystyle{u_{\rm flow}}}{\displaystyle{10\ {\rm km\ s}^{-1}}}\biggr)\biggl(\frac{\displaystyle{l_{\rm flow}}}{\displaystyle{1\ {\rm pc}}}\biggr)^{-1}
\end{equation} 
and
\begin{equation}
E_{\rm flow}\sim L_{\rm flow}t_{\rm flow}\sim 4.6\times 10^{47}{\rm erg\ s}^{-1}\biggl(\frac{\displaystyle{L_{\rm flow}}}{\displaystyle{40\ \LO}}\biggr)\biggl(\frac{\displaystyle{t_{\rm flow}}}{\displaystyle{10^5\ {\rm yr}}}\biggr),
\end{equation}
where $u_{\rm flow}$, $l_{\rm flow}$, and $t_{\rm flow}$ are 
the typical velocity, size, 
and lifetime of the outflow lobe, respectively, and $L_{\rm flow}$ represents
the typical mechanical luminosity of the outflow lobes in a molecular cloud. 
When we take the collision frequency,
\begin{equation}
\gamma\rho_{\rm i}\sim 1.7\times 10^{-13}\mbox{s}^{-1}\biggl(\frac{\displaystyle{n_{\rm i}/n}}{\displaystyle{10^{-7}}}\biggr)\biggl(\frac{\displaystyle{n}}{\displaystyle{10^3\ \mbox{cm}^{-3}}}\biggr),
\end{equation}
as the normalization of the frequency, the frequency of the outflow lobes
can written as $\omega_{\rm flow}\sim$ 2 $\gamma\rho_{\rm i}$.

Next, we estimate the energy of waves generated from the outflows.
Kato (1968) studied the generation of Alfv\'{e}n waves from a turbulent
source in a fully ionized, magnetized plasma. He showed that the efficiency 
of the energy inputs from the source to the waves is almost unity if 
the Alfv\'{e}n Mach number of the source is larger than unity.
Since in the outflow lobes the velocity ($\sim$ 10 km s$^{-1}$) seems
to be super Alfv\'{e}nic, its entire energy will turn into the waves.
Thus, $E_0$ used in section 3 corresponds to $E_{\rm flow}$ in this case.
The ratio $E_{\rm wave}/E_{\rm flow}$ at $t=10^5$ yr with the source of
the frequency of 2 $\gamma\rho_{\rm i}$ is calculated as $\sim$ 0.73
(see also figure 2).
Therefore, the energy, 
\begin{equation}
E_{\rm wave}\sim 0.73\ E_{\rm flow}\sim 3.4\times 10^{47} {\rm erg}, 
\end{equation}
propagates as waves, and the remaining fraction,
\begin{equation}
E_{\rm flow}-E_{\rm wave}\sim 1.2\times 10^{47} {\rm erg}, 
\end{equation}
dissipates and heats the neutrals due to ion--neutral damping.

At last we compare the timescale of the energy injection from the outflows,
$t_{\rm inj}$, with the timescale of the wave dissipation 
via the ion--neutral damping, $t_{\rm diss}^{\rm wave}$. 
Because the injection timescale is comparable to the lifetime of
an outflow, $t_{\rm flow}$, we obtain
\begin{equation}
t_{\rm inj}\sim 1\times10^5 {\rm yr}.
\end{equation}
Because the dissipation timescale corresponds to the collision time, 
$(\gamma\rho_i)^{-1}$, 
\begin{equation}
t_{\rm diss}^{\rm wave}\sim 2\times 10^{5}\mbox{yr}\biggl(\frac{\displaystyle{n_i/n}}{\displaystyle{10^{-7}}}\biggr)^{-1}\biggl(\frac{\displaystyle{n}}{\displaystyle{10^3\mbox{cm}^{-3}}}\biggr)^{-1}.
\end{equation}
{}From equations (30) and (31),
we find that $t_{\rm inj}$ is slightly shorter than 
$t_{\rm diss}^{\rm wave}$; 
that is, $t_{\rm inj}\ltsim t_{\rm diss}^{\rm wave}$.
This means just the same as the above result; i.e., more than a half of
the source energy propagates as waves.

In the following, we compare the above-mentioned energy and timescales 
with others in molecular clouds
to consider a significance of outflows as an energetic source 
in molecular clouds.

\subsection{Energetic Source of Interstellar Turbulence}

Here, we consider the energy inputs from outflows 
to turbulence in a molecular cloud. 
The absorption or scattering of waves propagating in a non-magnetized
turbulent medium have been investigated both theoretically and experimentally
(Hunter, Lowson 1974; Noir, George 1978; Salikuddin et al.\ 1988). 
In a partially ionized, 
magnetized medium, hydromagnetic waves may be absorbed in the same manner.
We thus compare the energy of the outflows with that of the turbulence 
in a molecular cloud to examine the possibility for the activity
of the outflows supporting the turbulence.
The energy of turbulence in a molecular cloud, $E_{\rm turb}$ is calculated as
\begin{equation}
E_{\rm turb}\sim \frac{\displaystyle{1}}{\displaystyle{2}}M_{\rm cloud}\sigma_{\rm turb}^2\sim 1\times 10^{47}\mbox{erg}\biggl(\frac{\displaystyle{M_{\rm cloud}}}{\displaystyle{10^4\ \MO}}\biggr)\biggl(\frac{\displaystyle{\sigma_{\rm turb}}}{\displaystyle{1\ {\rm km\ s}^{-1}}}\biggr)^2,
\end{equation}
where $M_{\rm cloud}$ and $\sigma_{\rm turb}$ are the typical mass of
a molecular cloud and the turbulent velocity at the typical size of 
the molecular cloud, respectively.
{}From this and equation (28), we find that the energy 
of interstellar turbulence 
is comparable to the energy of hydromagnetic waves in a molecular cloud,
namely, $E_{\rm turb}\sim E_{\rm wave}$.

By recent numerical simulations, the dissipation timescale 
of interstellar turbulence were calculated as
\begin{equation}
t_{\rm diss}^{\rm turb}\sim (0.25 \mbox{--} 2)\times 10^6\mbox{yr}
\end{equation}
(Stone et al.\ 1998). Equations (30) and (33) suggest
$t_{\rm inj}\ltsim t_{\rm diss}^{\rm turb}$, 
which means that the energy inputs
from the outflows to the interstellar turbulence can balance with the 
dissipation of the turbulence, which occurs mainly due to the shocks.
Thus, the turbulence may be in equilibrium in molecular clouds.
In any case, the outflows play an important role for energy inputs
to the turbulence in molecular clouds. 

\subsection{Heating Source in Molecular Clouds}

Here, we confirm that the dissipation of waves generated from outflows
plays an important role in molecular clouds.
A variety of heating sources have been considered in molecular clouds, 
such as cosmic rays, H$_2$ formation on grains,
gas--grain collisions, the compression or collapse of dense cores, 
and dissipation of turbulence, etc. (Spitzer 1978; Hollenbach 1988). 
The heating rate due to the cosmic rays, which is common as a heating
source in molecular clouds, is 
\begin{equation}
\Gamma_{\rm CR}\sim 6.4\times 10^{-25} \mbox{erg cm}^{-3}\mbox{s}^{-1}\biggl(\frac{\displaystyle{n}}{\displaystyle{10^3\ {\rm cm}^{-3}}}\biggr).
\end{equation}
Stone, Ostriker, and Gammie (1998) have estimated the heating rate 
based on the dissipation
of interstellar turbulence, making use of numerical simulations, as 
\begin{equation}
\Gamma_{\rm turb}\sim 5.8\times 10^{-25}\mbox{erg\ cm}^{-3}\mbox{s}^{-1}\biggl(\frac{\displaystyle{n}}{\displaystyle{10^3\ {\rm cm}^{-3}}}\biggr)\biggl(\frac{\displaystyle{\sigma_{\rm turb}}}{\displaystyle{1\ {\rm km\ s}^{-1}}}\biggr)^3\biggl(\frac{\displaystyle{l_{\rm cloud}}}{\displaystyle{10\ {\rm pc}}}\biggr)^{-1},
\end{equation}
where $l_{\rm cloud}$ is the typical size of a molecular cloud.
We now calculate the heating rate via the dissipation of the waves 
generated by the outflows as
\begin{equation}
\Gamma_{\rm wave}\sim 1.5\times 10^{-24}\mbox{erg cm}^{-3}\mbox{s}^{-1}\biggl(\frac{\displaystyle{E_{\rm flow}-E_{\rm wave}}}{\displaystyle{1.2\times 10^{47}\ \mbox{erg}}}\biggr)\biggl(\frac{\displaystyle{t_{\rm flow}}}{\displaystyle{10^5\ \mbox{yr}}}\biggr)^{-1}\biggl(\frac{\displaystyle{l_{\rm cloud}}}{\displaystyle{10\ \mbox{pc}}}\biggr)^{-3}.
\end{equation}
This is slightly larger than the heating rate of the other heating sources.
We note that this heating rate is comparable to the cooling rate,
due to optically thin CO line emission,
\begin{equation}
\Lambda_{\rm cool}\sim 4.1\times 10^{-24}\mbox{erg cm}^{-3}\mbox{s}^{-1}\biggl(\frac{\displaystyle{T}}{\displaystyle{10\ {\rm K}}}\biggr)^{2.2},
\end{equation}
where $T$ is the temperature in the molecular clouds 
(Goldsmith, Langer 1978).
We thus conclude that dissipated waves originating from outflows is
one of the heating sources in molecular clouds.

\section{Summary}

We investigated the evolution of hydromagnetic waves
generated by a compact turbulent source in partially ionized, 
magnetized plasmas, by applying the Lighthill theory.
We calculated how much of the energy of
the generated waves is dissipated to the ambient molecular gas
via ion--neutral damping, depending on the frequency 
of the oscillating source.
As a result, we found that the ratio of the dissipated energy to the energy of 
the propagating waves becomes larger as the frequency 
becomes higher, and also increases with passing time. The new point 
of our paper that is superior to Balsara (1996) is that we treat 
the wave equations with source terms as Lighthill did; namely,
we consider the generation of waves from a source as well as 
their propagation.

The implications to the phenomena in molecular clouds are also presented.
We discuss the effects of outflows as an energy source in molecular clouds
when they act as compact turbulent sources. 
Our conclusions are: 

1. Since the timescale of the ion--neutral 
damping is slightly longer than the typical lifetime of the outflows,
more than half of the energy generated by the outflows
propagates as waves, and the remainder dissipates in the
molecular gas via ion--neutral damping.

2. The energy of the propagating waves is comparable to that of 
the turbulence in molecular clouds.
Interestingly, the dissipation timescale of the turbulence is
comparable to the lifetime of the outflows,
suggesting that it is possible for the hydromagnetic waves 
generated by the outflows to support the turbulence in molecular clouds.
Although our discussion partly confirms the suggestions by
Myers and Lazarian (1998),
the essential difference between ours and the previous studies is 
that we consider the evolution of waves originating from the outflows.

3. The heating rate due to outflows through the ion--neutral damping
of the generated waves is slightly larger than those of the other well-known 
heating mechanisms in molecular clouds. That is to say, outflow 
is one of the dominant heating sources which sustain the temperature 
of the molecular clouds. 

To conclude, we confirm that outflows generally play a substantial
role as an energy source in molecular clouds,
in the context of the Lighthill theory. Although we have applied our formula
to only simple cases for the first step in this paper, 
it could be developed to contribute to studies on the global evolution
of molecular clouds.\par

\vspace{1pc}\par

We would like to thank the referee, Prof. S.W. Stahler, for his careful
reading of this manuscript and valuable comments, which improved the
clarity of our discussions.

\setcounter{equation}{0}
\renewcommand{\theequation}{A\arabic{equation}}

\section*{Appendix.\ Derivation of Equations (12) and (13)}

Making use of equation (9),
the basic equations are written as follows: the continuity equation,
\begin{equation}
\frac{\displaystyle{\partial\rho_1}}{\displaystyle{\partial t}}+\rho_0\biggl(\frac{\displaystyle{\partial v_{1x}}}{\displaystyle{\partial x}}+\frac{\displaystyle{\partial v_{1y}}}{\displaystyle{\partial y}}+\frac{\displaystyle{\partial v_{1z}}}{\displaystyle{\partial z}}\biggr)=0;
\end{equation}
the momentum equation with equations (3),(7), and (8),
\begin{equation}
\frac{\displaystyle{\partial v_{1x}}}{\displaystyle{\partial t}}+[(\mbox{\boldmath $v$}_1\cdot\nabla)\mbox{\boldmath $v$}_1]_x=-\frac{\displaystyle{c_{\rm s}^2}}{\displaystyle{\rho_0}}\frac{\displaystyle{\partial\rho_1}}{\displaystyle{\partial x}}+[(\nabla\times\mbox{\boldmath $h$})\times\mbox{\boldmath $h$}]_x,
\end{equation}
\begin{equation}
\frac{\displaystyle{\partial v_{1y}}}{\displaystyle{\partial t}}+[(\mbox{\boldmath $v$}_1\cdot\nabla)\mbox{\boldmath $v$}_1]_y=-\frac{\displaystyle{c_{\rm s}^2}}{\displaystyle{\rho_0}}\frac{\displaystyle{\partial\rho_1}}{\displaystyle{\partial y}}+\frac{\displaystyle{B_0}}{\displaystyle{4\pi\rho_0}}\biggl(\frac{\displaystyle{\partial B_{1y}}}{\displaystyle{\partial x}}-\frac{\displaystyle{\partial B_{1x}}}{\displaystyle{\partial y}}\biggr)+[(\nabla\times\mbox{\boldmath $h$})\times\mbox{\boldmath $h$}]_y,
\end{equation}
and
\begin{equation}
\frac{\displaystyle{\partial v_{1z}}}{\displaystyle{\partial t}}+[(\mbox{\boldmath $v$}_1\cdot\nabla)\mbox{\boldmath $v$}_1]_z=-\frac{\displaystyle{c_{\rm s}^2}}{\displaystyle{\rho_0}}\frac{\displaystyle{\partial\rho_1}}{\displaystyle{\partial z}}+\frac{\displaystyle{B_0}}{\displaystyle{4\pi\rho_0}}\biggl(\frac{\displaystyle{\partial B_{1y}}}{\displaystyle{\partial x}}-\frac{\displaystyle{\partial B_{1x}}}{\displaystyle{\partial y}}\biggr)+[(\nabla\times\mbox{\boldmath $h$})\times\mbox{\boldmath $h$}]_z,
\end{equation}
where $\mbox{\boldmath $h$}\equiv\mbox{\boldmath $B$}_1/(4\pi\rho_0)^{1/2}$,
and the induction equation with equations (6) and (8),
\begin{equation}
\frac{\displaystyle{\partial B_{1x}}}{\displaystyle{\partial t}}-B_0\biggl(\frac{\displaystyle{\partial v_{1y}}}{\displaystyle{\partial y}}+\frac{\displaystyle{\partial v_{1z}}}{\displaystyle{\partial z}}\biggr)+[\nabla\times(\mbox{\boldmath $B$}_1\times\mbox{\boldmath $v$}_1)]_x=\frac{\displaystyle{c_{\rm A}^2}}{\displaystyle{\gamma\rho_{\rm i}}}\nabla^2B_{1x},
\end{equation}
\begin{equation}
\frac{\displaystyle{\partial B_{1y}}}{\displaystyle{\partial t}}-B_0\frac{\displaystyle{\partial v_{1y}}}{\displaystyle{\partial x}}+[\nabla\times(\mbox{\boldmath $B$}_1\times\mbox{\boldmath $v$}_1)]_y=\frac{\displaystyle{c_{\rm A}^2}}{\displaystyle{\gamma\rho_{\rm i}}}\frac{\displaystyle{\partial}}{\displaystyle{\partial x}}\biggl(\frac{\displaystyle{\partial B_{1y}}}{\displaystyle{\partial x}}-\frac{\displaystyle{\partial B_{1x}}}{\displaystyle{\partial z}}\biggr),
\end{equation}
and
\begin{equation}
\frac{\displaystyle{\partial B_{1z}}}{\displaystyle{\partial t}}-B_0\frac{\displaystyle{\partial v_{1z}}}{\displaystyle{\partial x}}+[\nabla\times(\mbox{\boldmath $B$}_1\times\mbox{\boldmath $v$}_1)]_z=\frac{\displaystyle{c_{\rm A}^2}}{\displaystyle{\gamma\rho_{\rm i}}}\frac{\displaystyle{\partial}}{\displaystyle{\partial x}}\biggl(\frac{\displaystyle{\partial B_{1z}}}{\displaystyle{\partial x}}-\frac{\displaystyle{\partial B_{1x}}}{\displaystyle{\partial y}}\biggr).
\end{equation}

At first, we derive equation (12). From equations (A3) and (A4), and
the variable,
\begin{equation}
\xi=\frac{\displaystyle{\partial v_{1z}}}{\displaystyle{\partial y}}-\frac{\displaystyle{\partial v_{1y}}}{\displaystyle{\partial z}},
\end{equation}
we obtain
\begin{equation}
\frac{\displaystyle{\partial^2\xi}}{\displaystyle{\partial t^2}}+\frac{\displaystyle{\partial}}{\displaystyle{\partial t}}\{\nabla\times[(\mbox{\boldmath $v$}_1\cdot\nabla)\mbox{\boldmath $v$}_1]\}_x=\frac{\displaystyle{B_0}}{\displaystyle{4\pi\rho_0}}\frac{\displaystyle{\partial}}{\displaystyle{\partial x}}\biggl(\frac{\displaystyle{\partial}}{\displaystyle{\partial z}}\frac{\displaystyle{\partial B_{1y}}}{\displaystyle{\partial t}}-\frac{\displaystyle{\partial}}{\displaystyle{\partial y}}\frac{\displaystyle{\partial B_{1z}}}{\displaystyle{\partial t}}\biggr)+\frac{\displaystyle{\partial}}{\displaystyle{\partial t}}\{\nabla\times[(\nabla\times\mbox{\boldmath $h$})\times\mbox{\boldmath $h$}]\}_x.
\end{equation}
By substituting equations (A6) and (A7), and using the assumption 
that the gas is fully ionized in the turbulent source region, 
the first term on the right-hand side of equation (A9) becomes
\begin{equation}
c_{\rm A}^2\frac{\displaystyle{\partial^2\xi}}{\displaystyle{\partial x^2}}+\frac{\displaystyle{c_{\rm A}^2}}{\displaystyle{\gamma\rho_{\rm i}}}\frac{\displaystyle{\partial^3\xi}}{\displaystyle{\partial x^2\partial t}}-c_{\rm A}\frac{\displaystyle{\partial}}{\displaystyle{\partial x}}\{\nabla\times[\nabla\times(\mbox{\boldmath $h$}\times\mbox{\boldmath $v$}_1)]\}_x.
\end{equation}
Thus, we can derive equation (12),
\begin{equation} 
\biggl[\frac{\displaystyle{\partial^2}}{\displaystyle{\partial t^2}}-\frac{\displaystyle{c_{\rm A}^2}}{\displaystyle{\gamma\rho_{\rm i}}}\frac{\displaystyle{\partial^3}}{\displaystyle{\partial t\partial x^2}}-c_{\rm A}^2\frac{\displaystyle{\partial^2}}{\displaystyle{\partial x^2}}\biggr]\xi(x,t) = [\nabla\times\mbox{\boldmath $q$}]_x,
\end{equation}
where
\begin{equation}
\mbox{\boldmath $q$}(\mbox{\boldmath $r$},t)=-\frac{\displaystyle{\partial}}{\displaystyle{\partial t}}[(\mbox{\boldmath $v$}_1\cdot\nabla)\mbox{\boldmath $v$}_1-(\nabla\times\mbox{\boldmath $h$})\times\mbox{\boldmath $h$}]-c_{\rm A}\{\nabla\times[\nabla\times(\mbox{\boldmath $v$}_1\times\mbox{\boldmath $h$}_1)]\}\times\mbox{\boldmath $e$}_x.
\end{equation}
Here, $\mbox{\boldmath $e$}_x=(1,0,0)$ is the direction of the uniform magnetic fields.

Next, we derive equation (13). From equations (A1)--(A4), and the variable
\begin{equation}
\eta=\frac{\displaystyle{\partial v_{1x}}}{\displaystyle{\partial x}}+\frac{\displaystyle{\partial v_{1y}}}{\displaystyle{\partial y}}+\frac{\displaystyle{\partial v_{1z}}}{\displaystyle{\partial z}},
\end{equation}
we obtain with equation (6),
\begin{equation}
\frac{\displaystyle{\partial^2\eta}}{\displaystyle{\partial t^2}}+\frac{\displaystyle{\partial}}{\displaystyle{\partial t}}\{\nabla\cdot[(\mbox{\boldmath $v$}_1\cdot\nabla)\mbox{\boldmath $v$}_1]\}=c_{\rm s}^2\nabla^2\eta+\frac{\displaystyle{B_0}}{\displaystyle{4\pi\rho_0}}\frac{\displaystyle{\partial}}{\displaystyle{\partial t}}\nabla^2B_{1x}+\frac{\displaystyle{\partial}}{\displaystyle{\partial t}}\{\nabla\cdot[(\nabla\times\mbox{\boldmath $h$})\times\mbox{\boldmath $h$}]\}.
\end{equation}
By substituting equation (A5), the second term in the right hand side of 
equation (A14) becomes
\begin{equation}
c_{\rm A}^2\nabla^2(\eta-\zeta)+\frac{\displaystyle{c_{\rm A}^2}}{\displaystyle{\gamma\rho_{\rm i}}}\frac{\displaystyle{B_0}}{\displaystyle{4\pi\rho_0}}\nabla^2B_{1x}-c_{\rm A}\nabla^2[\nabla\times(\mbox{\boldmath $h$}\times\mbox{\boldmath $v$}_1)]_x,
\end{equation}
where $\zeta\equiv\partial v_{1x}/\partial x$. Now, from equations (A1) and
(A2) we obtain
\begin{equation}
\frac{\displaystyle{\partial^2\zeta}}{\displaystyle{\partial t^2}}+\frac{\displaystyle{\partial}}{\displaystyle{\partial t}}\frac{\displaystyle{\partial}}{\displaystyle{\partial x}}[(\mbox{\boldmath $v$}_1\cdot\nabla)\mbox{\boldmath $v$}_1]_x=c_{\rm s}^2\frac{\displaystyle{\partial^2\eta}}{\displaystyle{\partial x^2}}-\frac{\displaystyle{\partial}}{\displaystyle{\partial t}}\frac{\displaystyle{\partial}}{\displaystyle{\partial x}}[(\nabla\times\mbox{\boldmath $h$})\times\mbox{\boldmath $h$}]_x.
\end{equation}
Thus, making use of equations (A14)--(A16) and the fully ionized condition
in the source region, we can derive equation (13) and wave equation for 
$\zeta$,
\begin{equation} 
\biggl[\frac{\displaystyle{\partial^4}}{\displaystyle{\partial t^4}}-\frac{\displaystyle{c_{\rm A}^2}}{\displaystyle{\gamma\rho_{\rm i}}}\nabla^2\frac{\displaystyle{\partial^3}}{\displaystyle{\partial t^3}}-(c_{\rm A}^2+c_{\rm s}^2)\nabla^2\frac{\displaystyle{\partial^2}}{\displaystyle{\partial t^2}}+\frac{\displaystyle{c_{\rm A}^2c_{\rm s}^2}}{\displaystyle{\gamma\rho_{\rm i}}}\nabla^4\frac{\displaystyle{\partial}}{\displaystyle{\partial t}}+c_{\rm A}^2c_{\rm s}^2\nabla^2\frac{\displaystyle{\partial^2}}{\displaystyle{\partial x^2}}\biggr]\eta(\mbox{\boldmath $r$},t) = Q_{\eta}(\mbox{\boldmath $r$},t),
\end{equation}
and
\begin{equation} 
\biggl[\frac{\displaystyle{\partial^4}}{\displaystyle{\partial t^4}}-\frac{\displaystyle{c_{\rm A}^2}}{\displaystyle{\gamma\rho_{\rm i}}}\nabla^2\frac{\displaystyle{\partial^3}}{\displaystyle{\partial t^3}}-(c_{\rm A}^2+c_{\rm s}^2)\nabla^2\frac{\displaystyle{\partial^2}}{\displaystyle{\partial t^2}}+\frac{\displaystyle{c_{\rm A}^2c_{\rm s}^2}}{\displaystyle{\gamma\rho_{\rm i}}}\nabla^4\frac{\displaystyle{\partial}}{\displaystyle{\partial t}}+c_{\rm A}^2c_{\rm s}^2\nabla^2\frac{\displaystyle{\partial^2}}{\displaystyle{\partial x^2}}\biggr]\zeta(\mbox{\boldmath $r$},t) = Q_{\zeta}(\mbox{\boldmath $r$},t),
\end{equation}
where
\begin{equation} 
Q_{\eta}(\mbox{\boldmath $r$},t)=\frac{\displaystyle{\partial^2}}{\displaystyle{\partial t^2}}(\nabla\cdot\mbox{\boldmath $q$})-c_{\rm A}^2\nabla^2\frac{\displaystyle{\partial q_x}}{\displaystyle{\partial x}},
\end{equation}
and
\begin{equation} 
Q_{\zeta}(\mbox{\boldmath $r$},t)=\frac{\displaystyle{\partial^2}}{\displaystyle{\partial t^2}}(\nabla\cdot\mbox{\boldmath $q$})-\biggl[\frac{\displaystyle{\partial^2}}{\displaystyle{\partial t^2}}-(c_{\rm A}^2+c_{\rm s}^2)\nabla^2\biggr]\frac{\displaystyle{\partial q_x}}{\displaystyle{\partial x}}.
\end{equation}

\clearpage
\section*{References} 
\re
Balsara D.S.\ 1996, ApJ 465, 775
\re
Bontemps S., Andr\'{e} P., Terebey S., Cabrit S.\ 1997, A\&A 311, 858
\re
Caselli P., Myers P.C.\ 1995, ApJ 446, 665
\re
Falgarone E., Puget J.--L., P\'{e}rault M.\ 1992, A\&A 257, 715
\re
Fleck R.C.\ Jr 1980, ApJ 242, 1019
\re
Franco J., Carrami\~{n}ana A.\ 1998, Interstellar Turbulence
(Cambridge Univ.\ Press, Cambridge) p1
\re
Fukui Y., Iwata T., Mizuno A., Bally J., Lane A.P.\ 1993,
in Protostars and Planets III, ed E.H.\ Levy, J.\ Lunine
(Univ. Arizona Press, Tuscon) p603
\re
Gammie C.F., Ostriker E.C.\ 1996, ApJ 466, 814
\re
Goldsmith P.F., Langer W.D.\ 1978, ApJ 222, 881
\re
Hollenbach D.\ 1988, Astrophys. Lett. 26, 191
\re
Hunter G.H., Lowson M.V.\ 1974, J.\ Acoust.\ Soc.\ Am.\ 55, 937
\re
Kamaya H., Nishi R.\ 1998, ApJ 500, 257
\re
Kato S.\ 1968, PASJ 20, 59
\re
Kudoh T., Shibata K.\ 1997, ApJ 474, 362 
\re
Kulsrud R.M.\ 1955, ApJ 121, 461
\re
Kulsrud R., Pearce W.P.\ 1969, ApJ 156, 445
\re
Larson R.B.\ 1981, MNRAS 194, 809
\re
Lee J.W.\ 1993, ApJ 404, 372
\re
Lighthill M.J.\ 1952, Proc.\ R.\ Soc.\ London A 211, 564
\re
Lighthill M.J.\ 1960, Phil.\ Trans.\ Roy.\ Soc.\ A 252, 397
\re
Lighthill M.J.\ 1978, Waves in Fluids (Cambridge Univ.\ Press, Cambridge) ch4.9
\re
Lizano S., Shu F.H.\ 1987, in Physical Processes in Interstellar Clouds,
ed G.E.\ Morfill, M.\ Scholer (Reidel, Dordrecht) p173
\re
Miesch M.S., Bally J.\ 1994, ApJ 429, 645
\re
Morse P.M., Feshbach H.\ 1953, Methods of Theoretical Physics 
(McGraw-Hill, New York) ch7
\re
Mouschovias T.C., Paleologou E.V.\ 1980, Moon Planets 22, 31
\re
Myers P.C., Benson P.J.\ 1983, ApJ 266, 309
\re
Myers P.C., Lazarian A.\ 1998, ApJ 507, L157
\re
Nakano T.\ 1998, ApJ 494, 587 
\re
Neufeld D.A., Lepp S., Melnick G.J.\ 1995, ApJS 100, 132
\re
Noir D.T., George A.R.\ 1978, J. Fluid Mech.\ 86, 593
\re
Nomura H., Kamaya H., Mineshige S.\ 1998, ApJ 508, 714
\re
Norman C., Silk J.\ 1980, ApJ 238, 158
\re
Salikuddin M., Tam C.K.W., Burrin, R.H.\ 1988, JSV 127, 91
\re
Scalo J.H.\ 1987, in Interstellar Processes, ed D.J.\ Hollenbach,
H.A.\ Thoronson Jr (Reidel, Dordrecht) p349
\re
Shu F.H.\ 1983, ApJ 273, 202
\re
Shu F., Najita J., Galli D., Ostriker E., Lizano S.\ 1993,
in Protostars and Planets III, ed E.H.\ Levy, J.\ Lunine
(Tuscon, Univ. Arizona Press) p3
\re
Spitzer L.\ 1978, Physical Processes in the Interstellar Medium 
(Wiley, New York) ch6
\re
Stahler S.W.\ 1994, ApJ 422, 616
\re
Stone J.M., Ostriker E.C., Gammie C.F. 1998, ApJ 508, L99
\re
Tomisaka K.\ 1998, ApJ 502, L163
\re
V\'{a}zquez-Semadeni E., Ballesteros-Paredes J., Rodr\'{\i}guez L.F.\ 1997, 
ApJ 474, 292
\re
Zweibel E.G., Josafatsson K.\ 1983, ApJ 270, 511 
\re
Zweibel E.G., McKee C.F.\ 1995, ApJ 439, 779

\clearpage
\centerline{Figure Captions}
\bigskip
\begin{fv}{1}
{}
{Real parts of (a) the $x$-component of perturbed vorticity, $\xi(x)$, 
and (b) the divergence of perturbed velocity, $\eta(r)$,
generated by a compact source with frequency of 
$\omega_0/\gamma\rho_{\rm i}=$ 0.5 (dashed lines), 1.0 (solid lines), 
and 5.0 (dotted lines). $\gamma\rho_{\rm i}$ represents the collision
frequency, and the compact source is located at $x=0$.}
\end{fv}
\begin{fv}{2}
{}
{Time evolution of the wave energy generated by a source with a frequency of 
$\omega_0/\gamma\rho_{\rm i}=$ 10$^{-8}$ (bold solid line),
0.5 (dashed line), 1.0 (solid line), 2.0 (dot-dashed line), 
and 5.0 (dotted lines).
Clearly, when the oscillation frequency is larger than 
the friction frequency, $\gamma\rho_{\rm i}$, the wave energy dissipates soon
in the molecular gas via ion--neutral damping.}
\end{fv}
\begin{fv}{3}
{}
{Imaginary parts of the wavenumber, $k_{\rm I}$,
as a function of the source frequency, $\omega_0$.
The units of the frequency is friction frequency, $\gamma\rho_{\rm i}$.
$k_{\rm I}$ for three types of the hydromagnetic waves, that is,
the Alfv\'{e}n mode (solid lines), the fast mode (dashed lines), 
and the slow mode (dotted lines) are represented.
When the frequency of the source is large, the generated waves easily damp,
because their amplitudes are proportional to $\exp(-|k_{\rm I}|x)$.}
\end{fv}
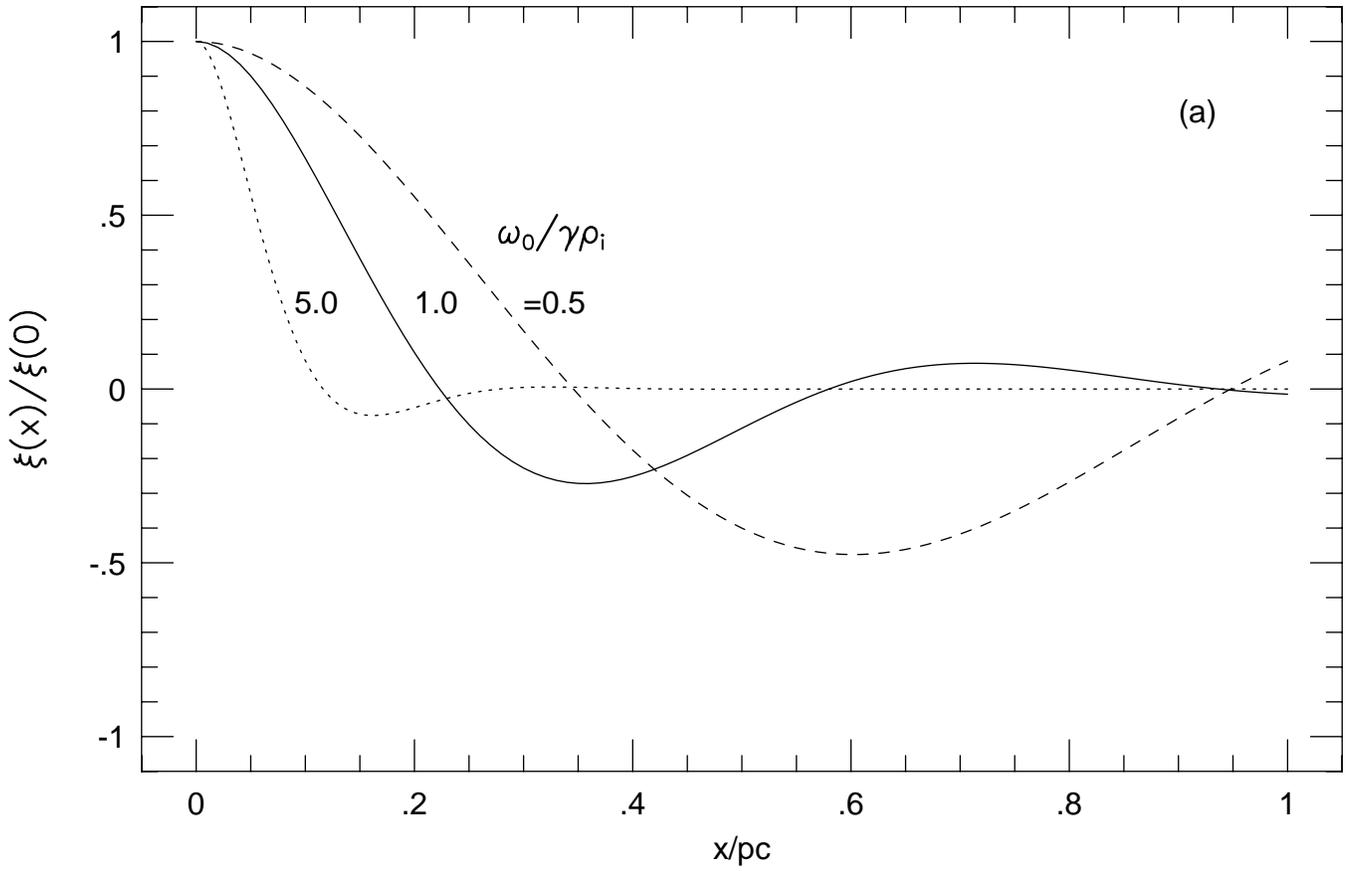
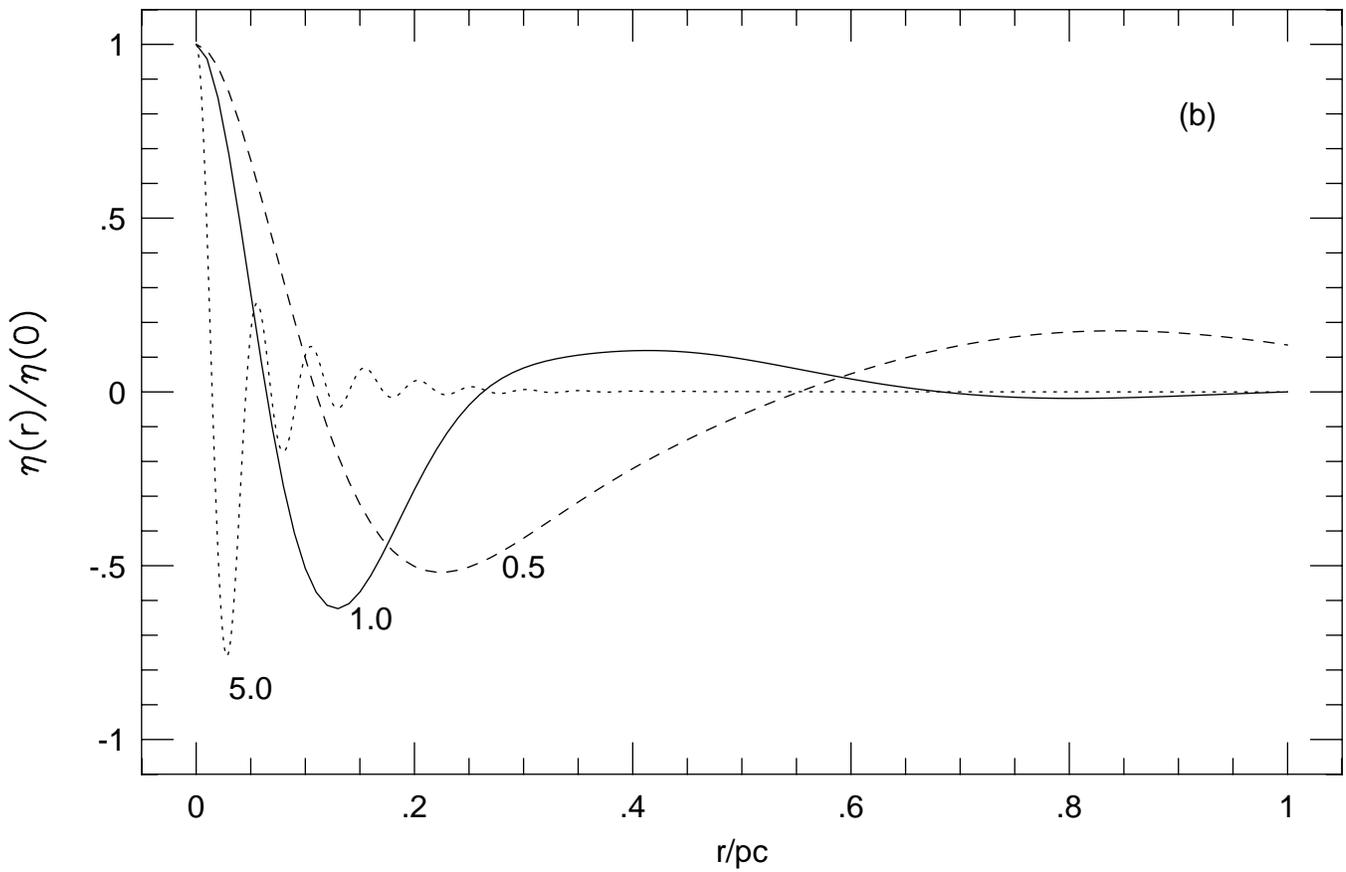
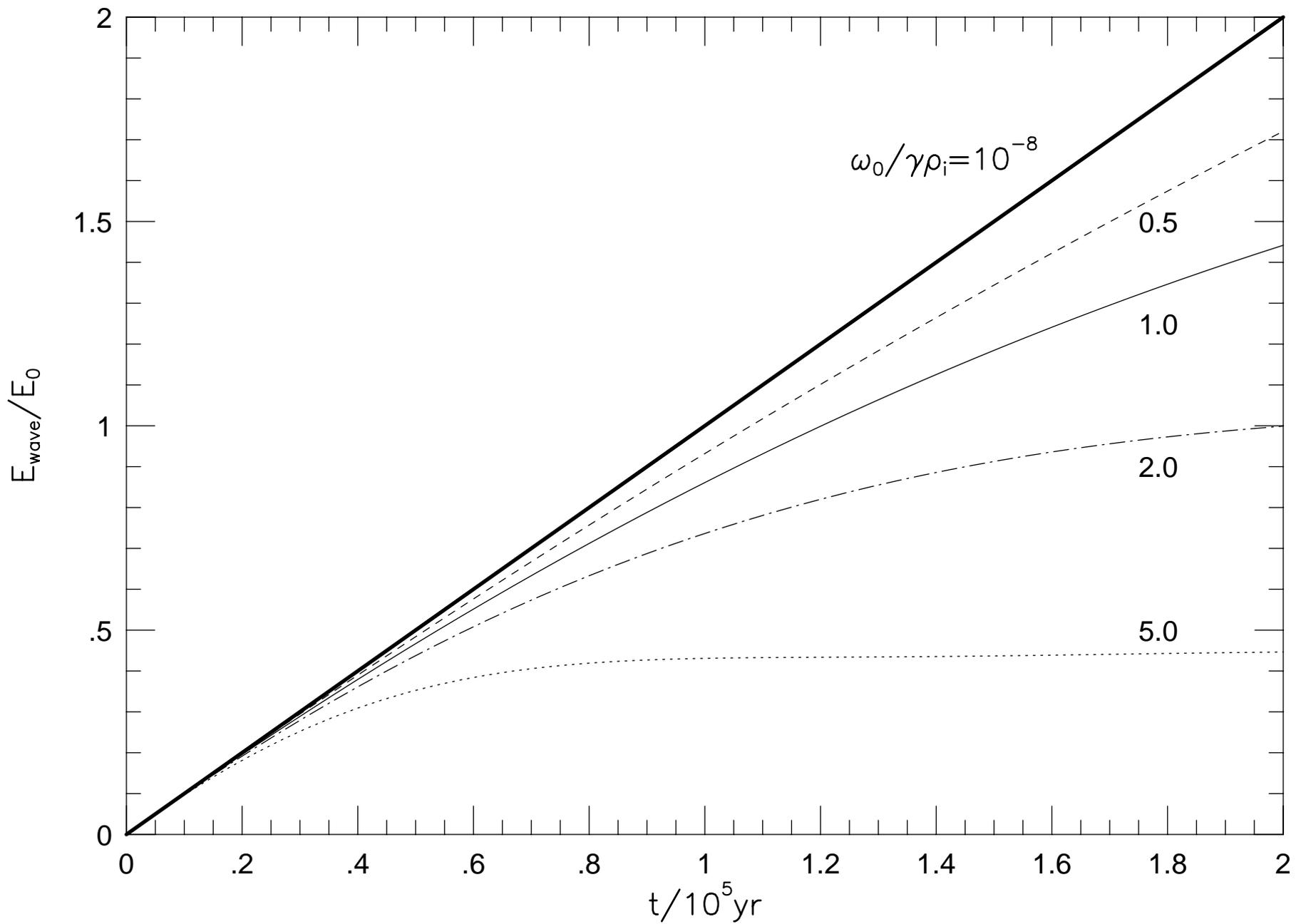
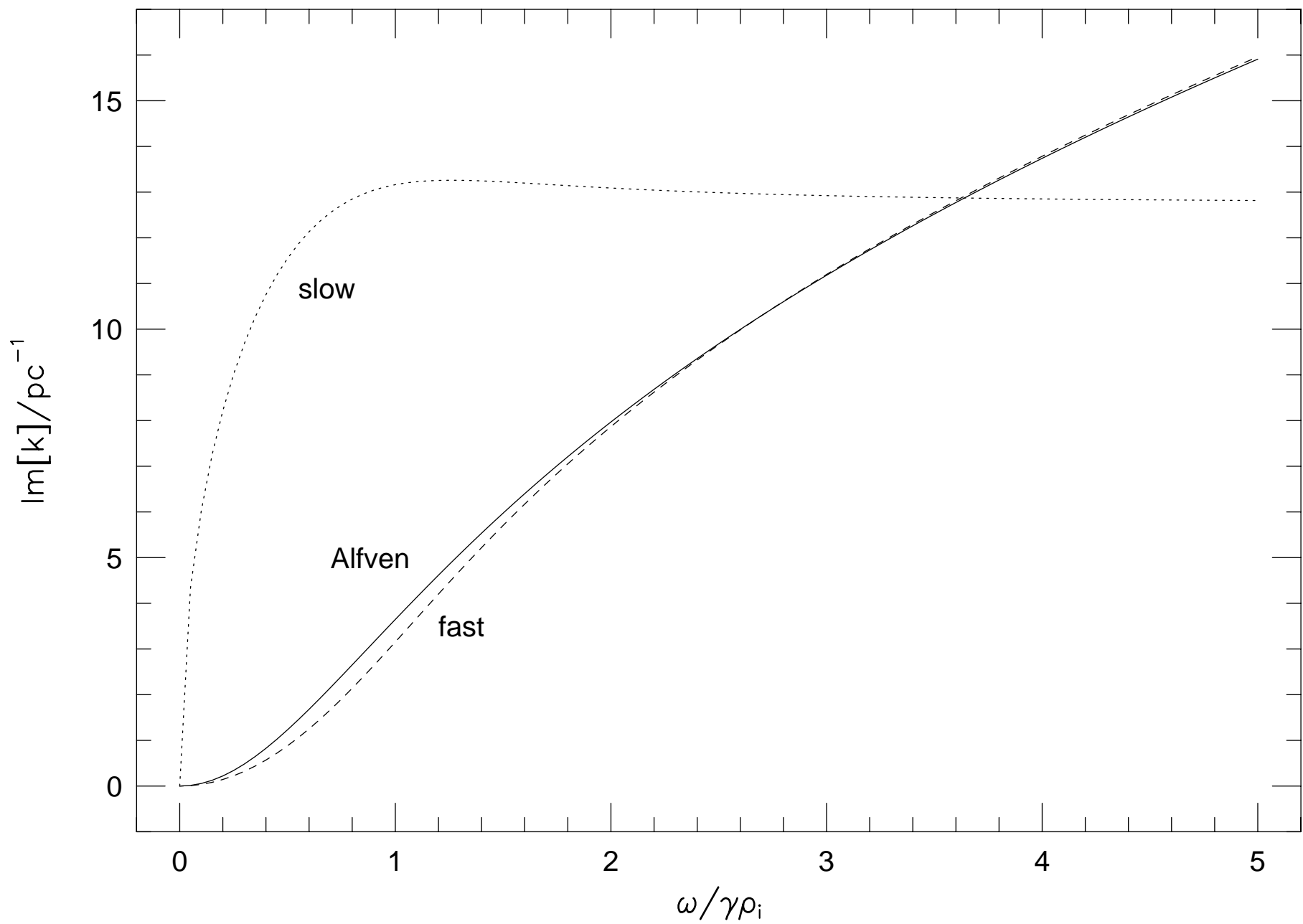
\end{document}